\documentclass[11pt,a4paper]{article}
\usepackage{jheppub,epsfig,multicol,bbm,latexsym,graphicx,subfigure}

\def\L{\left(}
\def\R{\right)}

\def\mdm{m_{\mbox{\tiny DM}}}
\def\sv{\ensuremath{\langle\sigma v\rangle}}
\def\f{\frac}
\title{Interpretation of the cosmic ray positron and electron excesses
with an annihilating-decaying dark matter scenario}

\author{Lei Feng$^{a}$, Zhaofeng Kang$^b$, Qiang Yuan$^{a,c,d}$,
Peng-Fei Yin$^{e}$, and Yi-Zhong Fan$^{a,c}$}

\affiliation{
$^a$Key Laboratory of Dark Matter and Space Astronomy, Purple Mountain Observatory, Chinese Academy of Sciences, Nanjing 210008, China\\
$^b$School of physics, Huazhong University of Science and Technology, Wuhan 430074, China\\
$^c$School of Astronomy and Space Science, University of Science and
Technology of China, Hefei 230026, China\\
$^d$Center for High Energy Physics, Peking University, Beijing 100871, China\\
$^e$Key Laboratory of Particle Astrophysics, Institute of High Energy Physics, Chinese Academy of Sciences, Beijing 100049, China
}
\emailAdd{fenglei@pmo.ac.cn}
\emailAdd{zhaofengkang@gmail.com}
\emailAdd{yuanq@pmo.ac.cn}
\emailAdd{yinpf@ihep.ac.cn}
\emailAdd{yzfan@pmo.ac.cn}

\abstract{
The precise measurements of energy spectra of cosmic ray positrons
and/or electrons by recent experiments show clear excesses above 10 GeV.
Moreover, a potential sharp spectral feature was suggested by the
Dark Matter Particle Explorer (DAMPE) data. These results inspire quite
a number of discussions on the connection with either the
annihilation/decay of dark matter (DM) or the astrophysical origins.
Here we discuss a DM scenario in which DM particles could
annihilate and decay into standard model particle pairs simultaneously.
In this model, the peak structure is due to the DM annihilation in a nearby
subhalo and the broad positron/electron excesses are due to the decay of
DM in the Milky Way. This model can reasonably explain the DAMPE and AMS-02
data of the total $e^+e^-$ spectra and the positron fraction, with
model parameters being consistent with existing constraints.
A simple realization of such a DM model is the spin-1 vector DM model.
}


\arxivnumber{1902.09235 }
\begin{document}
\maketitle
\flushbottom

\section{Introduction}

Dark matter (DM) particles may annihilate or/and decay into standard model
particles such as pairs of electrons/positrons and protons/anti-protons,
and can hence give rise to excesses in the cosmic ray spectra. Identification
of such excesses is one of the most important goals of the indirect detection
experiments of DM such as PAMELA~\cite{Adriani:2008zr},
ATIC~\cite{Chang:2008aa}, Fermi-LAT~\cite{Abdo:2009zk,Abdollahi:2017nat},
and AMS-02~\cite{Accardo:2014lma,Aguilar:2014fea}.
Some progresses have been made in recent years. The most widely-known
phenomena are perhaps the spectral anomalies (excesses) of positrons and the
electron plus positron spectra. Either astrophysical source(s)/process(es)
\cite{Yuksel:2008rf,Hooper:2008kg,Profumo:2008ms,Malyshev:2009tw,Hu:2009bc,
Blasi:2009hv,Linden:2013mqa,Yin:2013vaa} or the DM annihilation/decay
\cite{Bergstrom:2008gr,Barger:2008su,Cirelli:2008pk,Yin:2008bs,Zhang:2008tb,Sergio:2010,
Feng:2013zca,Blum:2013zsa,Bergstrom:2013jra,Kopp:2013eka,Cholis:2013psa,
Yuan:2013eja,Jin:2013nta,Dev:2013,Belotsky:2015}) were proposed to account for the data.

The DArk Matter Particle Explorer (DAMPE, \cite{Chang:2014,TheDAMPE:2017dtc}),
launched on December 17, 2015, is a high-energy particle detector dedicated
to DM indirect detection and cosmic ray physics.
%It has great energy resolution which is better than $\rm 1.5\% @TeV$ for
%electrons and gamma rays. It also has good hadron rejection power (higher
%than $10^5$) that can detect the cosmic ray electrons and gamma-rays with
%excellent energy resolution and very low background contamination.
The DAMPE collaboration reported the precise measurement of the cosmic
ray $e^+ + e^-$ (CREs) spectrum from 25 GeV to 4.6 TeV \cite{Ambrosi:2017wek}.
The DAMPE data confirms the spectral hardening of CREs around 50~GeV
\cite{Abdollahi:2017nat}, and reveals clearly a spectral softening at
$\sim 0.9$~TeV \cite{Ambrosi:2017wek}. These results are consistent with
previously reported CRE excesses \cite{Chang:2008aa,Abdo:2009zk,
Abdollahi:2017nat,Aguilar:2014fea}. Moreover, there might be a sharp peak
at $\sim 1.4 $ TeV. Although the significance of the current data is
relatively low ($\sim2.3\sigma$ considering the look-elsewhere effect
\cite{Fowlie:2017fya}), this structure, if confirmed, should imply the
existence of nearby quasi-monoenergetic electron sources~\cite{Yuan:2017ysv}.

If the DM annihilation or decay is employed to account for the peak, the
DM particles should annihilate or decay dominantly into leptons, and the
annihilation or decay should occur in local regions not far away from the
solar system. This is because that TeV CREs lose their energies very quickly
when travelling in the Milky Way, and can not reach us if they were
generated far away. The quark channels should be dramatically suppressed
due to the constraints of anti-protons. Such models have been proposed in
literature \cite{Jin:2017qcv,Zu:2017dzm,Fan:2017sor,Gu:2017gle,Duan:2017pkq,
Tang:2017lfb,Chao:2017yjg,Gu:2017bdw,Athron:2017drj,Cao:2017ydw,Duan:2017qwj,
Liu:2017rgs,Huang:2017egk,Gao:2017pym,Niu:2017hqe,Chen:2017tva,Li:2017tmd,
Zhu:2017tvk,Gu:2017lir,Nomura:2017ohi,Ghorbani:2017cey,Cao:2017sju,
Yang:2017cjm,Ding:2017jdr,Liu:2017obm,Ge:2017tkd,Zhao:2017nrt,Sui:2017qra,
Okada:2017pgr,Chao:2017emq,Cao:2017rjr,Han:2017ars,Niu:2017lts,
Nomura:2018jkd,Wang:2018pcc,Pan:2018lhc,Liu:2019iik}.
See Ref.~\cite{Yuan:2018rys} for a special review of relevant studies.

The main purpose of this work is to interpret both the sub-TeV positron
and electron excesses and the DAMPE peak simultaneously in an annihilating
plus decaying DM (ADDM) model. It is natural to speculate that
DM particles with a limited lifetime can at the same time annihilate
with each other. While the annihilation products, presumed to be leptons,
can explain the peak excess around 1.4 TeV, the decay products with
lower energies can account for the sub-TeV electron/positron excesses.
We will investigate whether such a scenario can be realized to account for
the data, without violating existing constraints from e.g., $\gamma$-ray
observations \cite{Chen:2009uq,Hutsi:2010ai,Cirelli:2012ut,Cheng:2016slx}
and the cosmic microwave background (CMB) data \cite{Ade:2015xua}.

This paper is organized as follows. In Section II, we compare ADDM model
prediction with the AMS-02 positron fraction and the AMS-02/DAMPE CRE data.
In Section III, we discuss a possible theoretical realization of
such an ADDM scenario. Our conclusions are summarized in Section IV.

\section{CREs from ADDM}

\subsection{Cosmic ray propagation}

The general form of cosmic ray propagation equation reads~\cite{Strong:2007nh}:
\begin{eqnarray}
\frac{\partial\psi}{\partial
t}&=&q({\bf r},p)+\nabla\cdot\left(D_{xx}\nabla\psi-{\bf V}\psi\right)+
\frac{\partial}{\partial p}p^{2}D_{pp}\frac{\partial}{\partial p}
\frac{\psi}{p^{2}} \nonumber \\
 &-& \frac{\partial}{\partial p}\left[\dot{p}\psi-\frac{p}{3}(\nabla\cdot
{\bf V})\psi\right]-\frac{\psi}{\tau_{f}}-\frac{\psi}{\tau_{r}},
\end{eqnarray}
where $\psi=\psi({\bf r},p,t)$ is the phase space density, $q({\bf r},p)$
denotes the source function, $D_{xx}=\beta D_0(E/4\,{\rm GeV})^{\delta}$
is the spatial diffusion coefficient, ${\bf V}=dV/dz\cdot {\bf z}$ represents
the convection velocity, $D_{pp}$ denotes the diffusion coefficient in the
momentum space which is characterized by the Alfvenic speed $v_A$ and is
to describe the reacceleration of particles in the interstellar medium,
$\dot{p}=dp/dt$ is the momentum loss rate, and $\tau_f$ ($\tau_r$) is the
time scale of the fragmentation (the radioactive decay).

Typically the Boron-to-Carbon ratio, the radioactive secondary
abundances, as well as diffuse $\gamma$-rays are used to constrain the
cosmic ray propagation parameters \cite{Strong:2007nh,Yuan:2017ozr,
Yuan:2018vgk,Ackermann:2012rg}. In this work, we consider a few propagation
model configurations and parameter settings, which are summarized in Table
\ref{tab1}. For the diffusion-reacceleration configuration, we also adopt
several values of the halo height following Ref.~\cite{Ackermann:2012pya}.
As a benchmark setting, we adopt the diffusion-reacceleration model with
a half-height of the propagation cynlinder $z_h=4$~kpc, as shown in bold in
Table \ref{tab1} (model II). Different propagation models affect mainly the
low-energy spectra of electrons and positrons, and have very minor effects
on the DM model which is the focus of this work. This is because the main
propagation effect of high-energy CREs is radiative cooling rather than
the diffusion, convection, and reacceleration. High-energy CREs can only
propagate a very limited distance in the Galaxy before they get cooled down.
As an illustration, the cooling time for TeV CREs is about $3\times 10^5$
years and the corresponding diffusion length is about 1 kpc for the diffusion
parameter setting II.

\begin{table}[!htb]
\centering
\caption {Propagation parameters. The benchmark setting is model II which
is shown in bold.}
\begin{tabular}{cccccc}
\hline \hline
Model & $D_0^a$ & $z_h$ & $v_A$ &$dV_c/dz$ & $\delta$   \\
      & $\rm (10^{28}~cm^2~s^{-1})$ & (kpc) & ($\rm km~s^{-1}$) & ($\rm km~s^{-1}~kpc^{-1}$) &  \\
\hline
I   & 2.7 & 2.0 & 35.0 & 0 &0.33   \\
II  & 5.3 & 4.0 & 33.5 & 0 &0.33  \\
III & 9.4 & 10.0 & 28.6 & 0 &0.33  \\
IV  & 4.2 & 10.9 & 0 & 5.4 &0.59  \\
\hline
\hline
\end{tabular}\\
$^a$Diffusion coefficient at rigidity of 4 GV.
\label{tab1}
\end{table}

This propagation equation can be solved numerically, by e.g.,
{\tt GALPROP}~\cite{Strong:1998pw} and {\tt DRAGON}~\cite{Evoli:2008dv}.
In this work we use the {\tt LikeDM} package~\cite{Huang:2016pxg}, which
tabulates the outputs from {\tt GALPROP} and enables fast computation,
to calculate the propagation of CREs for all the background (non-DM)
components and the Milky Way DM annihilation/decay component of CREs.
Note that we will also deal with the CRE propagation from a nearby DM
subhalo. For this particular case, we simplify the propagation equation,
keeping only the diffusion and energy loss terms, and adopt the analytical
Green's function to solve the CRE propagation in a spherically symmetric
geometry with infinite boundary conditions~\cite{Atoyan:1995ux}.

%For nearby CR source, the propagation of CR electrons/positrons is time dependent and their cooling rate in the
%local region can be approximated as \cite{Atoyan:1995ux}
%\begin{equation}
%-{\rm d}E/{\rm d}t\equiv b(E)=b_0+b_1E_{\rm GeV}+b_2E_{\rm GeV}^2,
%\label{be}
%\end{equation}
%where $b_{\rm 0} \approx {\rm 3\times10^{-16} GeV s^{-1}}$, $b_{\rm 1} \approx {\rm 10^{-15} GeV s^{-1}}$ and $b_{\rm 2} \approx {\rm 1.0 \times10^{-16} GeV s^{-1}}$. The cooling time of electrons/positrons is defined as $\tau(E)\equiv E/B(E)$, and the effective propagation length $\lambda(E)$ within its cooling time is
%\begin{equation}
%\lambda(E)=2\left(\int_E^{\infty}\frac{D(E')}{b(E')}dE'\right)^{1/2},
%\end{equation}
%where $D(E)$ denotes the spatial diffusion coefficient. The propagation
%of local high energy electrons can be solved analytically if we adopt a spherically
%symmetric geometry with infinite boundary conditions (see \cite{Atoyan:1995ux} for more details).

\subsection{DM distribution}

For the annihilation process, the CRE source function is
\begin{equation}
q(E,r)=\frac{\sv}{2\mdm^2}\frac{{\rm d}N}{{\rm d}E}\times \rho^2(r),
\end{equation}
where $\mdm$ is the mass of the DM particle and ${\rm d}N/{\rm d}E$ is
the spectrum of CREs per annihilation. For the decaying case, the source
term is
\begin{equation}
q(E,r)=\frac{\rho(r)}{\mdm\tau}\frac{{\rm d}N}{{\rm d}E},
\label{eq:2}
\end{equation}
where $\tau$ is the lifetime of the DM particle.

The DM distribution $\rho(r)$ in the Milky Way is assumed to be an isothermal
distribution~\cite{Bahcall:1980fb}
\begin{equation}
\rho_{\rm mw}(r)=\frac{\rho_s}{1+(r/r_s)^2},
\label{isothermal}
\end{equation}
where $\rho_s=1.16$~GeV~cm$^{-3}$ denotes the finite central density and
$r_s=5$~kpc represents the core radius. As for the subhalo, we assume an
NFW distribution~\cite{Navarro:1996gj}
\begin{equation}
\rho_{\rm sub}\left(r \right)=\frac{\delta_{c} \rho_{\rm crit}}
{\left( r/r_{\rm s}\right)\left(1+r/r_{\rm s}\right)^{2}},
\end{equation}
where $r_{s}$ and $\delta_{c}$ are the scale radius and characteristic
density, $\rho_{\rm crit} =3H^{2}/8 \pi G$ is the critical density of
the Universe. The $\delta_c$ parameter relates with the subhalo
concentration parameter $c=r_v/r_s$, where $r_v$ is the virial radius,
as~\cite{Springel:2008cc}
\begin{equation}
\delta_c=7.213\delta_V=\frac{200}{3}\frac{c^3}{\ln(1+c)-c/(1+c)}.
\end{equation}
For subhalos in the solar neighborhood, we have
approximately~\cite{Springel:2008cc}
\begin{equation}
\delta_V=1.2\times 10^{6}\left(\frac{M_{\rm sub}}{10^{6} ~ M_{\odot}}\right)^{-0.18},
\end{equation}
where $M_{\rm sub}$ is the mass of the subhalo. The tidal force of the
Milky Way DM halo would remove the DM beyond a so-called tidal radius
from the subhalo. Adopting the method of Ref.~\cite{Springel:2008cc},
the tidal radius is found to be roughly 0.2 times of the original virial
radius of a subhalo~\cite{Yuan:2017ysv}. Therefore, the DM distribution
of the subhalo is an NFW distribution truncated at the tidal radius $r_t$.

\subsection{Results}

%The data used in this study include the AMS-02 positron fraction
%\cite{Accardo:2014lma}, the AMS-02 CRE spectrum in the energy range
%of $\rm 0.5~GeV \sim 25~GeV$~\cite{Aguilar:2014fea}, and the DAMPE CRE
%spectrum~\cite{Ambrosi:2017wek}. The AMS-02 CRE fluxes above 25 GeV are
%not used because of a systematical difference from that of DAMPE.

The background electrons include primary electrons accelerated from
conventional cosmic ray sources, whose injection spectrum is
parameterized as a three-segment broken power-law with an exponential
cutoff~\cite{Yuan:2017ysv}, and secondary electrons from inelastic
collisions of cosmic ray nuclei and the interstellar medium.
The background electron spectral parameters are tuned to match
the data. The background positrons are mainly from the inelastic collisions
between cosmic ray nuclei and the medium. The detailed parameters of the CR electron and positron backgrounds are presented in Table \ref{tab3} in the Appendix.
The annihilation and decay contributions from DM in both the Milky Way halo and the nearby subhalo
are also added to the model.

\begin{figure}[htbp]
\centering
\includegraphics[width=0.48\columnwidth]{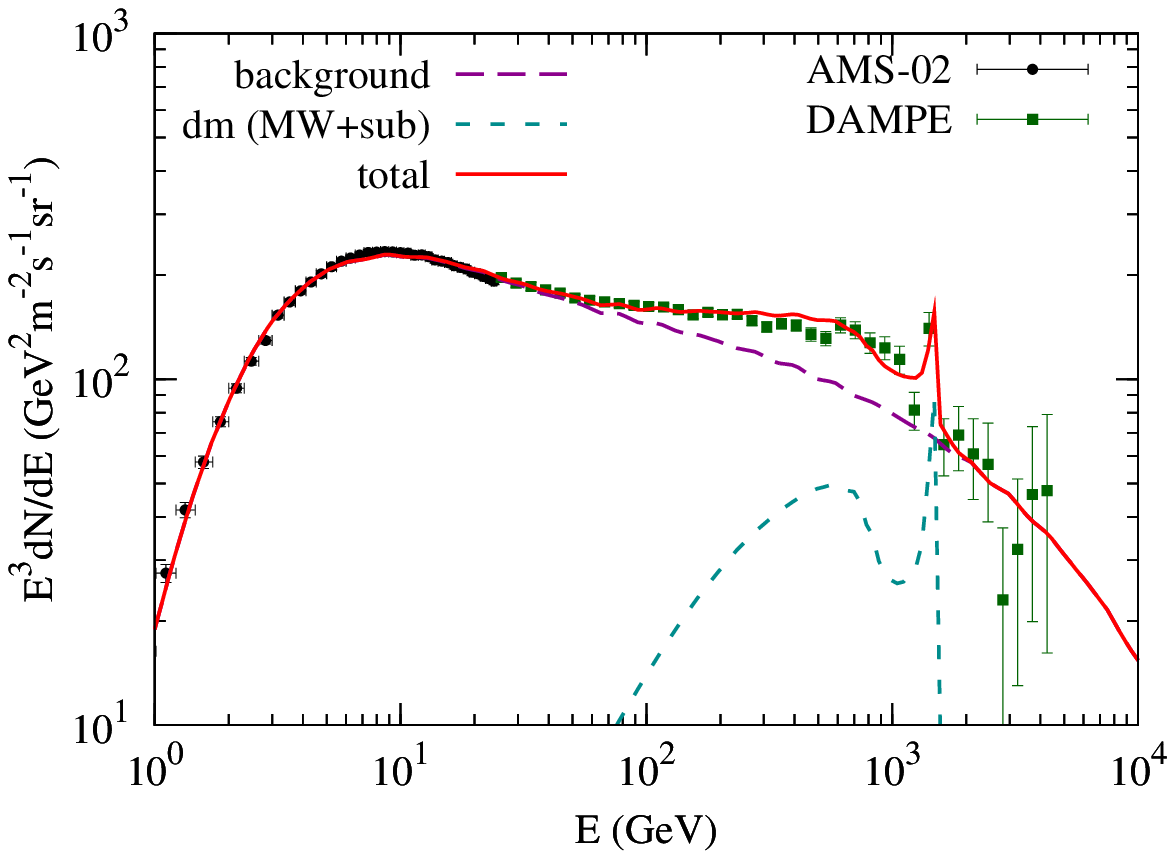}
\includegraphics[width=0.48\columnwidth]{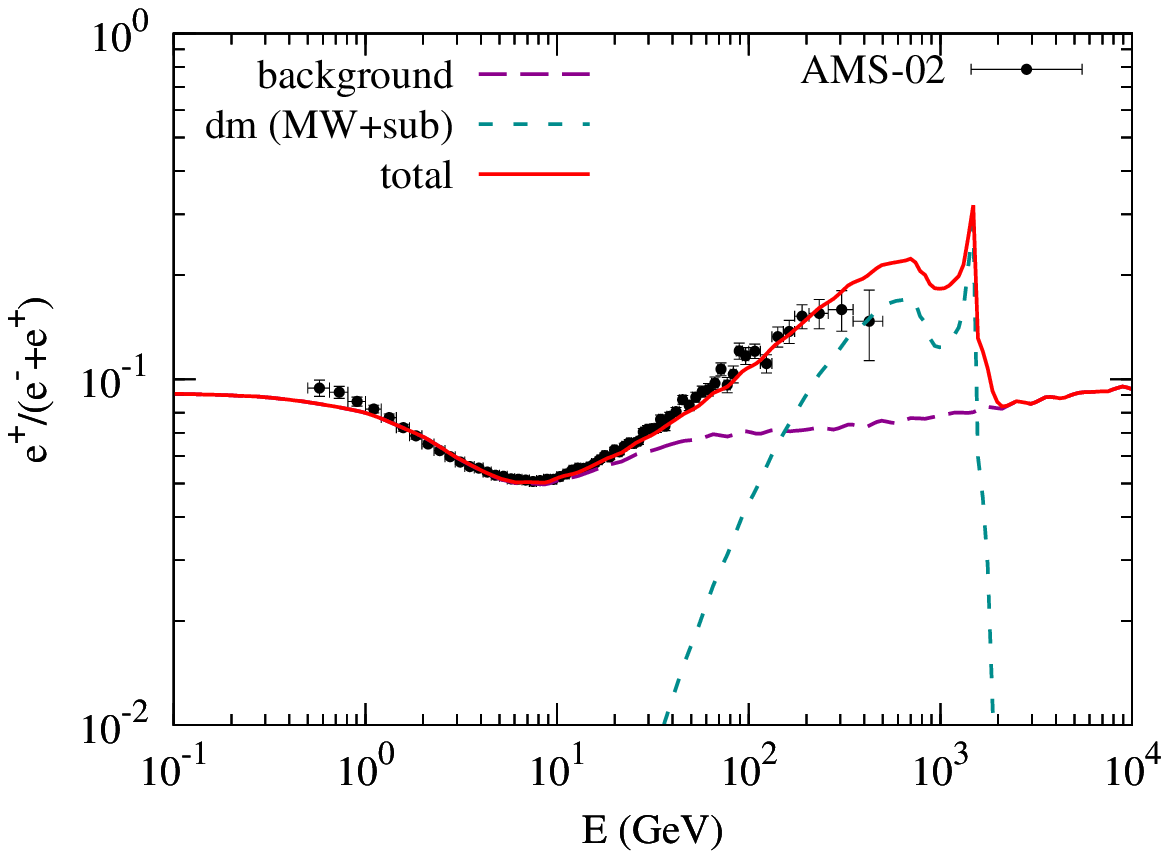}
\caption{The total CRE fluxes (left) and positron fraction (right) for the
ADDM model for propagation model II, compared with the AMS-02
\cite{Accardo:2014lma,Aguilar:2014fea} and DAMPE~\cite{Ambrosi:2017wek} data.
Here the DM particles annihilate and decay into $e^+e^-$ and $\mu^+\mu^-$
with branching ratios of $1:1$. Other model parameters are, $\mdm =
{\rm 1.5~TeV}$, $\sv={\rm 2.8\times10^{-24}~cm^3~s^{-1}}$,
$\tau=9\times10^{26}$~s, $M_{\rm sub}=5\times10^4$~M$_{\odot}$,
and $d=0.1$~kpc.}
\label{equalcontribution}
\end{figure}

\begin{figure}[htbp]
\centering
\includegraphics[width=0.48\columnwidth]{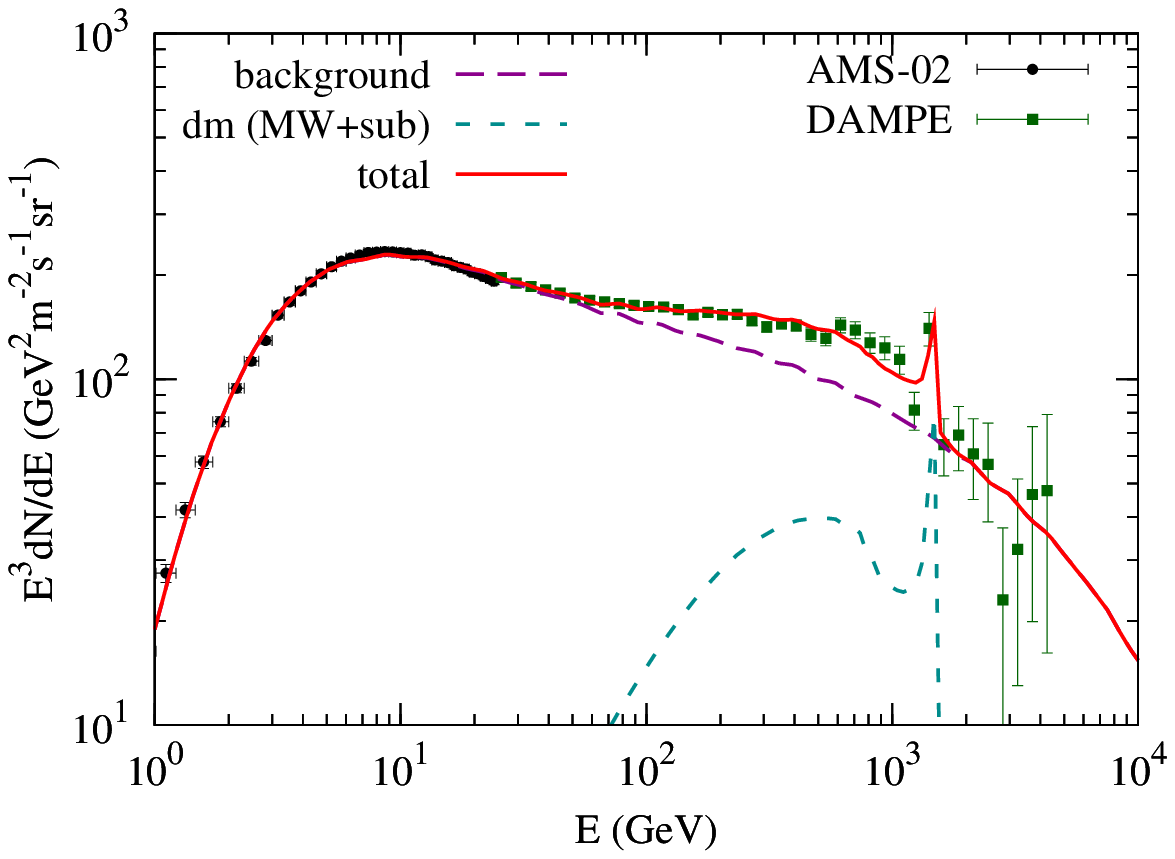}
\includegraphics[width=0.48\columnwidth]{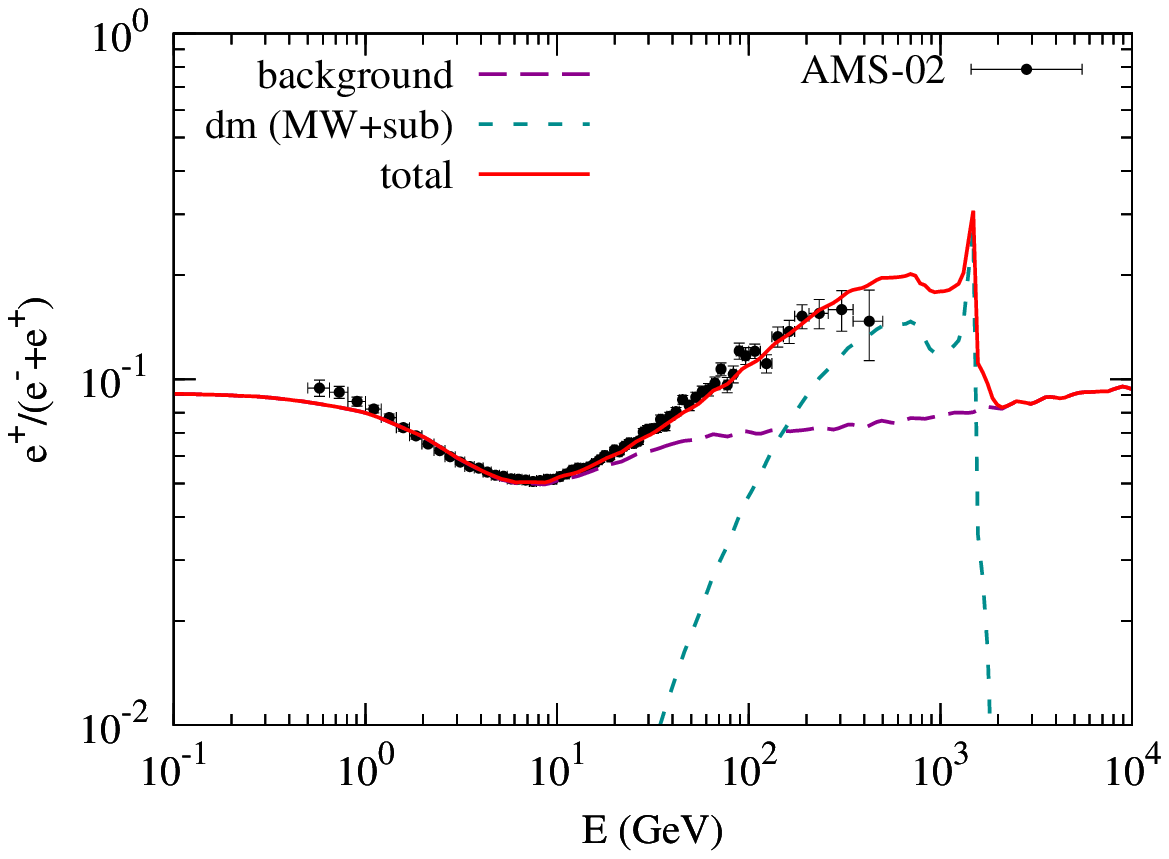}
\caption{Same as Fig.~\ref{equalcontribution} but for branching ratio of
$e:\mu=1:3$. The mass of subhalo is about $10^5$~M$_\odot$. The other
parameters are the same as those in Fig.~\ref{equalcontribution}.
}
\label{noequalcontribution}
\end{figure}

Here we take the propagation model setting II as an example to discuss the fitting results.
Figure \ref{equalcontribution} shows the CRE spectrum (left panel) and
the positron fraction (right panel) from the model prediction for the
propagation model setting II, compared with the measurements
\cite{Accardo:2014lma,Aguilar:2014fea,Ambrosi:2017wek}.
Here the ADDM model parameters are: the mass of the DM particle is
$\mdm = {\rm 1.5~TeV}$, the annihilation cross section is
$\sv={\rm 2.8\times10^{-24}~cm^3~s^{-1}}$, the decay lifetime is
$\tau=9\times10^{26}$~s, and the branching ratios are $e:\mu=1:1$.
The subhalo mass is $M_{\rm sub}=5\times10^4$~M$_{\odot}$, and the
distance to the subhalo center is $d=0.1$~kpc.
Figure \ref{noequalcontribution} shows a slightly improved fitting
with annihilation/decay branching ratios $e:\mu=1:3$.
It is shown that the model prediction matches well with the data.
The parameter values for the other propagation models are compiled
in Table \ref{tab2}, which show minor variations among different
models. For reference, the comparisons between the model and data
for the other three propagations models are given in Figure \ref{other}
in the Appendix.

\begin{table}[!htb]
\centering
\caption{The ADDM model parameters and required subhalo masses for different
propagation parameters. The DM annihilation/decay branching ratios are assumed to be $e:\mu=1:1$.}
\begin{tabular}{ccccc}
\hline \hline
Model & $\mdm$ & $\sv$ & $\tau$ & $M_{\rm sub}$ \\
 & (TeV) & ($10^{-24}$~cm$^3$~s$^{-1}$) & ($10^{26}$~s) & ($10^4$\,M$_{\odot}$) \\
\hline
I   & 1.5 & 2.8 & 8.1 & 3.5 \\
II  & 1.5 & 2.8 & 9.0 & 5.0 \\
III & 1.5 & 2.8 & 11.3 & 8.0 \\
IV  & 1.5 & 2.8 & 9.0 & 4.5 \\
\hline
\hline
\end{tabular}
\label{tab2}
\end{table}

\begin{figure}[htbp]
\centering
\includegraphics[width=0.7\columnwidth]{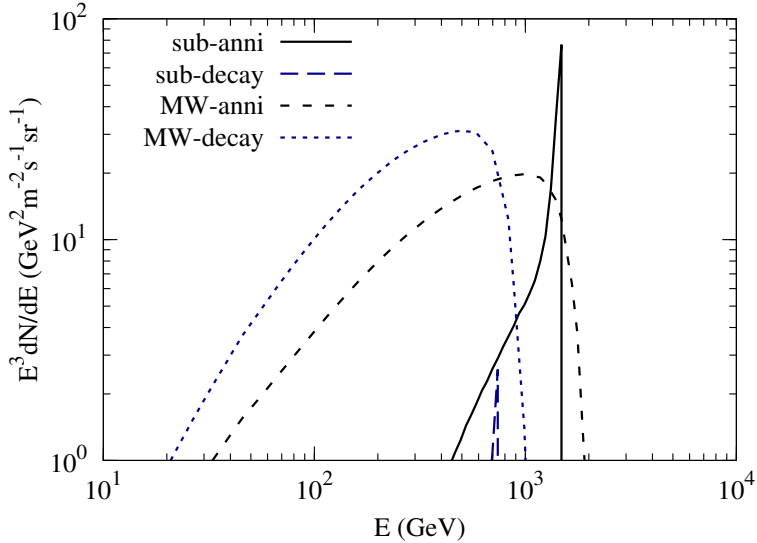}
\caption{Contributions of the subhalo and MW components to the CRE fluxes
for the annihilation and decay modes, respectively. The DM model parameters
are the same as those in Figure~\ref{equalcontribution}.}
\label{fig:MW-sub}
\end{figure}

The spike structure around 1.4 TeV is due to the annihilation of DM in
the subhalo, and the sub-TeV broad excesses of positrons and CREs are
mainly due to the decay of DM in the Milky Way halo. To clearly see this,
we plot in Figure \ref{fig:MW-sub} the CRE fluxes from the DM annihilation
or decay in either the Milky Way or the subhalo separately. For the
parameters we adopt, the subhalo contribution is dominated by the DM
annihilation, which gives rises to the 1.4 TeV peak shown in the data.
For the Milky Way components, the decay component is slightly larger
than the annihilation one. Note that here we need a relatively high
contribution from the DM annihilation in the Milky Way (and hence a
relatively large cross section), otherwise the CRE data from 700 GeV
to TeV cannot be well reproduced.

%At $\rm \sim 1.4~TeV$, there is a spike peak in the positron ratio data
%in this model. Such feature can be tested by the future experiments.

The annihilation cross section and decaying lifetime are marginally
consistent with the constraints from $\gamma$-rays \cite{Chen:2009uq,
Hutsi:2010ai,Cirelli:2012ut,Cheng:2016slx} and CMB \cite{Ade:2015xua}.
As shown in Ref.~\cite{Yuan:2017ysv}, the upper limits of the annihilation
cross section from CMB are about $3\times10^{-24}$~cm$^3$~s$^{-1}$ for
the $\mu^+\mu^-$ channel and $10^{-24}$~cm$^3$~s$^{-1}$ for the $e^+e^-$
channel. The lower limits on the decay lifetime from the extragalactic
diffuse $\gamma$-ray background are about $4\times10^{26}$~s for the
$\mu^+\mu^-$ channel and $10^{27}$~s for the $e^+e^-$ channel.
Therefore, the model parameters derived in this work are not excluded
by the current data.

The DM annihilation and decay in a nearby subhalo should also be
constrained by $\gamma$-rays and CRE anisotropies \cite{Yuan:2017ysv,
Coogan:2019}. In addition, the profile of the line-like feature can also
constrain the parameters of the subhalo \cite{Huang:2017egk,Coogan:2019}.
A large distance of subhalo ($\gtrsim0.3$~kpc) is ruled out by the width
of DAMPE CRE excess, and the halo size is bounded by the Fermi $\gamma$-ray
observations \cite{Coogan:2019}. In a forthcoming work, we will show that
the $\gamma$-ray emission from this postulated DM subhalo is consistent
with the current Fermi data at least in some low-latitude sky regions
\cite{Ge:2020}. Finally, we comment that the probablity of finding such
a massive subhalo within a short distance from the Earth is relatively
low \cite{Yuan:2017ysv,zhao:2019cpc}. According to Fig.~8 of
Ref.~\cite{Yuan:2017ysv}, we estimate that such a probability is about
$0.1\%$ for $d=0.1$~kpc and $M_{\rm sub}=5\times10^4$~M$_{\odot}$.

\section{Model}

According to the results in the previous section, we need a primary spectrum
which includes contributions from both annihilation and decay of DM into
a pair of light charged leptons (referring to $e$ and $\mu$, which are
commonly denoted as $\ell$ hereafter). The simplest candidate is supposed
to be a spin-0 particle $S$. However, this scenario does not work naturally.
We first consider that $S$ annihilates into $\bar\ell_i\ell_j$ via a
$t$-channel charged fermionic mediator. To avoid the $p$-wave suppression,
it requires a large chiral violation which at the same time gives rise to
either a large lepton flavor violation $\bar\ell_i\rightarrow\ell_j+\gamma$
or a large $g_\ell-2$; both have been ruled out. Then we consider the
$s$-channel annihilation exchanging a spin-0 or spin-1 mediator $X$.
The latter is again $p$-wave suppressed, whereas the former is not well
motivated.

\begin{figure}
\includegraphics[width=0.48\columnwidth]{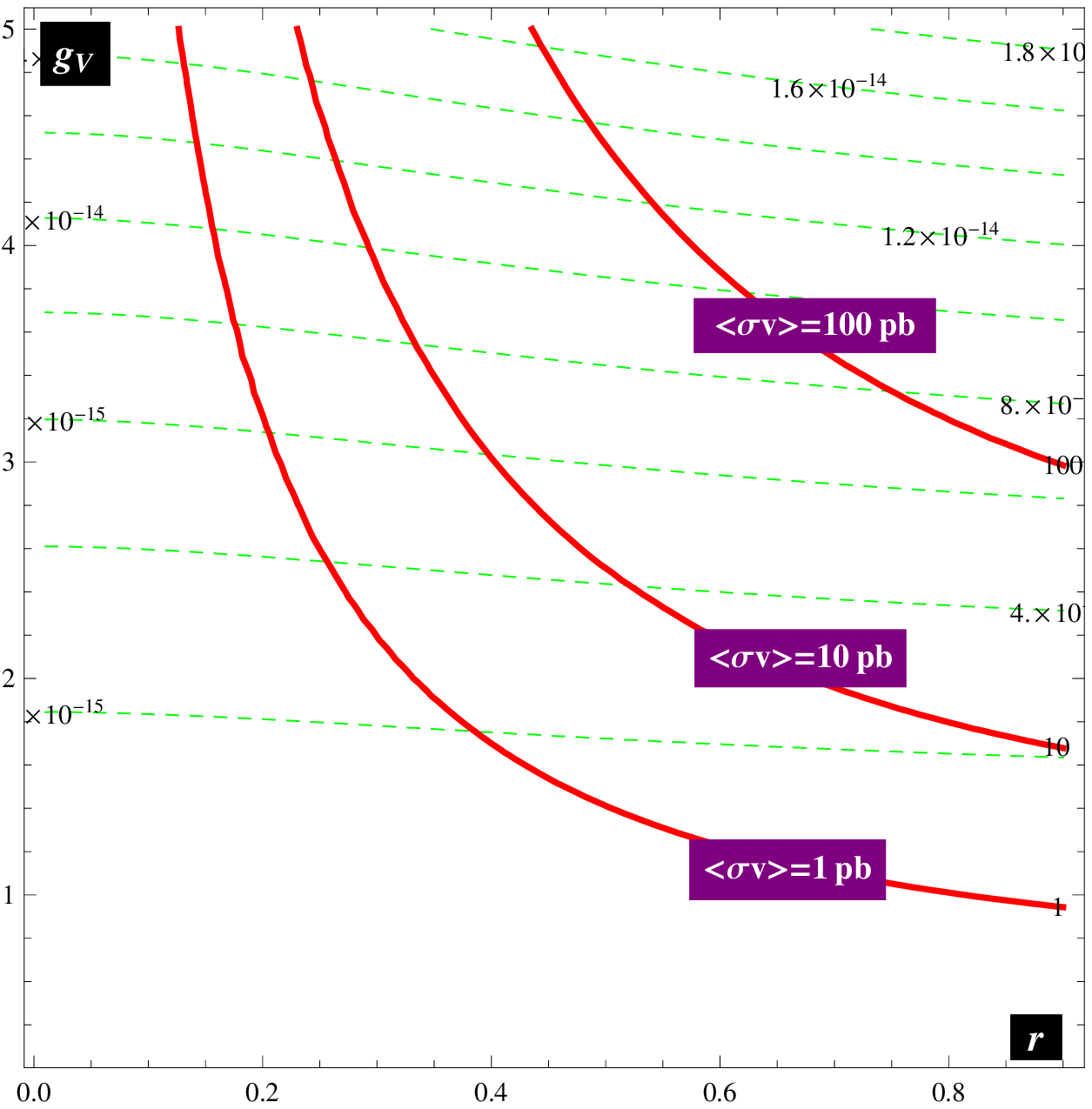}
\includegraphics[width=0.48\columnwidth]{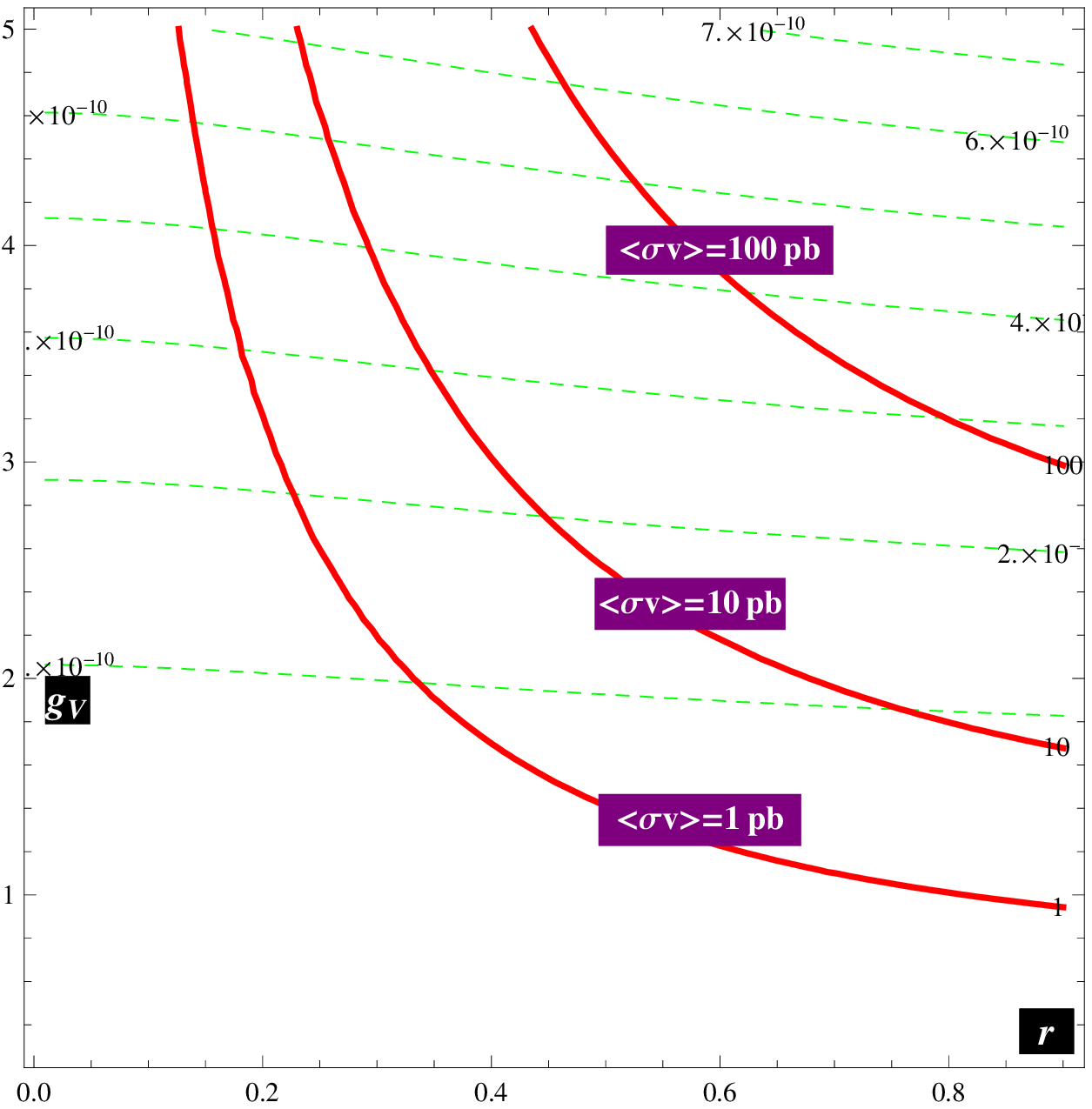}
\caption{Contours of the annihilation cross section times relative velocity
(red thick lines) and the contribution to $g_\ell-2$ (green dashed lines)
in the $r-g_V$ plane. The left panel shows the results for $g_e-2$, and
the right panel shows those for $g_\mu-2$. $m_V=1.5$ TeV is adopted.}
\label{gl-an}
\end{figure}

A spin-1 vector DM (VDM) $V_\mu$ may works. The reason is that its annihilation
into $\bar\ell\ell$ is not $p$-wave suppressed even in the absence of chiral
violation. In this work we are not aiming at constructing a complete model.
We just consider the following minimal $Z_2$-invariant effective model
\begin{equation}
-{\cal L}_1=\left(\frac{1}{2}m_VV^2+m_F\bar FF\right)+\left(g_V V_\mu\bar\ell \gamma^\mu P_L F+h.c.\right),
\label{VDM:an}
\end{equation}
where the $Z_2$-odd fermion $F$ is a Dirac fermion, mediating the $t$-channel
annihilation $VV\to \bar\ell \ell$. Here the VDM is leptophilic because it
couples dominantly to light leptons. To suppress the lepton flavor violation,
we further require that each lepton flavor $\ell_i$ has its own partner
$F_i$, and they do not give rise to new flavor violation. Different from the
scalar DM case, the annihilation cross section has unsuppressed $s$-wave
contribution
\begin{equation}
\langle \sigma_{\bar\ell\ell} v\rangle=\frac{g_V^4}{9\pi}\frac{1}{m_V^2}\frac{r^4}{(r^2+1)^2},
\label{anni}
\end{equation}
with $r\equiv m_V/m_F<1$. Without chiral violation, the contribution to
$g_\ell-2$ from Eq.~(\ref{VDM:an}) is not significant, given
by~\cite{Agrawal:2014ufa}
\begin{equation}
g_\ell-2=-|g_V|^2\frac{m_\ell^2}{m_V^2}\frac{5-14r^2+39r^4-38r^6+8r^8+18r^4\ln r^2}{ 96\pi^2 (1-r^2)^4},
\label{g-2}
\end{equation}
which is suppressed by the light fermion mass square. We display the numerical
results of Eq.~(\ref{anni}) and Eq.~(\ref{g-2}) in Fig.~\ref{gl-an}. We find
that for $\langle \sigma_{\bar\ell\ell} v\rangle$ as large as $\sim 100$ pb,
the resulting $g_\ell-2$ can still lie below the current uncertainties,
which are $\sim 10^{-12}$ and $10^{-9}$ for electrons and muons, respectively.
The conclusion is particularly true in the region with $r\sim 1$.  Note that for an annihilation cross section as large as 100 pb, the relic density after freezing-out will be too small to account for the total DM budget. One way to overcome this problem is assuming that the early universe after reheating does not reach a very high temperature, but much below the TeV scale. Consequently, the VDM and as well  the mediator $F$ never reach full thermal equilibrium, so one can not calculate VDM relic density via the usual freeze-out dynamics. Instead, one should track all productions of DM via the scattering like $\ell\bar\ell \rightarrow VV$, namely through freeze-in~\cite{Chen:2017kvz}

It is seen that $g_V\sim 4$ is needed to make $\langle \sigma_{\ell\bar \ell} v\rangle\sim 100$ pb,
so one should worry if such a VDM is subjected to a strong constraint by direct detection. Since this VDM is leptonic and moreover self-conjugate (a real vector), its leading interaction with the quark electromagnetic current is via the anapole momentum of VDM, described by the dimension-6 operator~\cite{Agrawal:2014ufa}
\begin{align}
{\cal O}_V=g_5 \f{e}{4\pi}\f{\alpha_V }{m_F^2}\partial_\mu V^\alpha \partial_\alpha V_\nu F_{\rho\sigma}\epsilon^{\mu\nu\rho\sigma},
\end{align}
with $F_{\rho\sigma}$ the Maxwell field strength tensor. The operator coefficient is estimated from the one-loop diagram with a charged loop, giving $\alpha_V=g_V^2/4\pi\sim 1$ and $g_5\sim {\cal O}(1)$. The resulting VDM-nucleon scattering is suppressed by $p$-wave in the nonrelativistic limit, which is explicit in the following estimation on the cross section~\footnote{We did not make a solid calculation, but borrowed the result from a similar model studied in Ref.~\cite{Ho:2012bg}, based on the Majorana fermion anapole dark matter. },
\begin{align}
\sigma_p\sim 0.5\times 10^{-13} \L\f{1.5 \rm TeV }{ m_F}\R^4\L\f{\alpha_V}{1.0}\R^2\L\f{v}{10^{-3}}\R^2 \rm pb.
\end{align}
It is four orders of magnitude below the current strongest bound from XENON1T. Actually, it already goes beneath the neutrino floor and therefore probably can not be probed in the direct detection experiments

Actually, a structure similar to Eq.~(\ref{VDM:an}) is presented in the
little Higgs model with $T$-parity~\cite{Birkedal:2006fz} or the extra
dimensional theory with KK-pairty~\cite{Servant:2002aq,Cheng:2002ej},
where the lightest $T/{\rm KK}$-odd vector boson is the VDM candidate.
In these models all left-handed fermions are accompanied with
$T/{\rm KK}$-odd heavy fermions. But after identifying $g_V$ with the
$U(1)_Y$ gauge coupling $g'$, the cross section is far from the required
value to explain the CRE data. From the required large $g_V$ value, it is
of interest to build VDM in the context of the composite Higgs model.

Now we discuss the VDM decay. To make it decay into a pair of light leptons,
the simplest approach is to introduce a term which breaks $Z_2$ explicitly
as the following:
\begin{equation}
-{\cal L}_{\rm decay}=\epsilon_V V_\mu\bar\ell \gamma^\mu \ell.
\label{VDM:de}
\end{equation}
The resulting lifetime of $V$ is $\tau_V\sim 10^{26}$~s~
$(10^{-26}/\epsilon_V)^2~(1.5~{\rm TeV}/m_V)$.
%The origin of the extremely small $\epsilon_V$ is beyond the scope of
%this paper. It needs completion of the effective model and as well
%specifying the origin of $Z_2$.
To understand the smallness of $\epsilon_V$, one needs the completion of the effective model and specifies the origin of $Z_2$. For instance, it might be due to a spontaneous but tiny break of $Z_2$ by a scalar field; such a break results in a tiny mixing between $V$ and some leptophilic gauge bosons, then giving rise to the effective operator in Eq.~(\ref{VDM:de}) through this tiny mixing. Alternatively, if $Z_2$ is identified with the $T$-parity, its violation may be due to anomaly~\cite{Hill:2007zv}.

To end up this section we make a comment on the spin-1/2 candidate $\chi$.
It may also give the desired CRE spectrum if its interaction is largely
specified by a spin-1 intermediate state $X_\mu$ which is merely slightly
lighter than $\chi$ and dominantly couples to $\ell$\footnote{The $t$-channel
annihilation again suffers from the $p$-wave suppression as the scalar
case~\cite{Baek:2015fma}.}. In such a case, the DM cascade decay
$\chi\to \nu+X(\to \bar \ell \ell)$ produces the $e^+e^-$ spectrum similar
to that from the two-body decay. Moreover, the annihilation
$\bar\chi+\chi\to \bar\ell_i+\ell_j$ is not $p$-wave suppressed. We leave
this scenario for future studies.

\section{Conclusion}\label{sec::con}

In this work, we propose an ADDM model to interpret the positron and CRE
data from AMS-02 and DAMPE. In this model we assume that DM particles
could annihilate and decay into light leptons (electrons and muons) at
the same time. It is shown that the annihilation in a nearby subhalo can
explain the potential peak structure of the CRE spectrum, and the decay
in the Milky Way halo can explain the broad sub-TeV excesses of both
positrons and CREs. The model parameters, although somehow tuned to explain
the data, are consistent with the current $\gamma$-ray and CMB constraints.

A spin-1 VDM was proposed as a particle realization of the model. The
annihilation into leptons of the VDM is not $p$-wave suppressed even in
the absence of chiral violation. At the same time, its contribution to
$g-2$ is not significant.

\begin{acknowledgments}
We thank Dr. Ran Ding for helpful discussions and suggestions.
This work was supported in part by the National Key Research and Development
Program of China (No. 2016YFA0400200), the National Natural Science Foundation
of China (Nos. 11773075, 11525313, 11722328, U1738210, U1738209, U1738206)
and the Youth Innovation Promotion Association of Chinese Academy of Sciences
(No. 2016288). QY is also supported by the 100 Talents Program of Chinese
Academy of Sciences.
\end{acknowledgments}

\section*{Appendix: results for the other cosmic ray propagation parameters}
The injection spectrum of the background electrons is parameterized
as a three-piece broken power-law with an exponential cutoff form, with
spectral indices $\nu_1$, $\nu_2$, $\nu_3$, break rigidities $R_{\rm br,1}$,
$R_{\rm br,2}$, and cutoff rigidity $R_{\rm cut}$. The background positron
spectrum is calculated according to the interaction between CR nuclei and
the interstellar medium \cite{Yuan:2017ozr}. A renormalization factor
($c_{e^+}$) has been multiplied to the secondary positron spectrum in order
to fit the low-energy positron fraction and CRE spectra simultaneously.
We further apply a force-field solar modulation model on the propagated
electron and positron spectra before comparing with the measurements
carried out on top of the atmosphere \cite{Gleeson:1968zza}.

\begin{table}[!htb]
\centering
\caption{The injection spectral parameters of the background electrons,
the renormalization factor of the background positrons, and the force-field
potential of the solar modulation.}
\begin{tabular}{ccccccccc}
\hline \hline
Model & $\nu_1$ & $\nu_2$ & $\nu_3$ & $R_{\rm br,1}$ & $R_{\rm br,2}$ & $R_{\rm cut}$ & $c_{e^+}$ & $\Phi$ \\
      &         &         &         & (GV)           & (GV)           & (GV)          &           & (GV) \\
\hline
I   & 1.48 & 3.11 & 2.61 & 2.2 & 41.0 & $4.0\times10^4$ & 2.72 & 1.83 \\
II  & 1.48 & 3.00 & 2.56 & 2.3 & 55.0 & $1.1\times10^4$ & 3.27 & 1.42 \\
III & 1.53 & 2.93 & 2.46 & 2.7 & 79.6 & $4.4\times10^3$ & 3.84 & 1.15 \\
IV  & 1.63 & 2.90 & 2.48 & 3.6 & 46.0 & $2.3\times10^4$ & 2.80 & 0.88 \\
\hline
\hline
\end{tabular}
\label{tab3}
\end{table}

%The parameter values for the other propagation models are compiled
%in Table \ref{tab2} for DM annihilation/decay branching ratios $e:\mu=1:1$.
%There are minor variations among different models.
%Heavier subhalo mass and longer decay lifetime are needed for larger half-height of the propagation cynlinder

\begin{figure}[htbp]
\centering
\includegraphics[width=0.48\columnwidth]{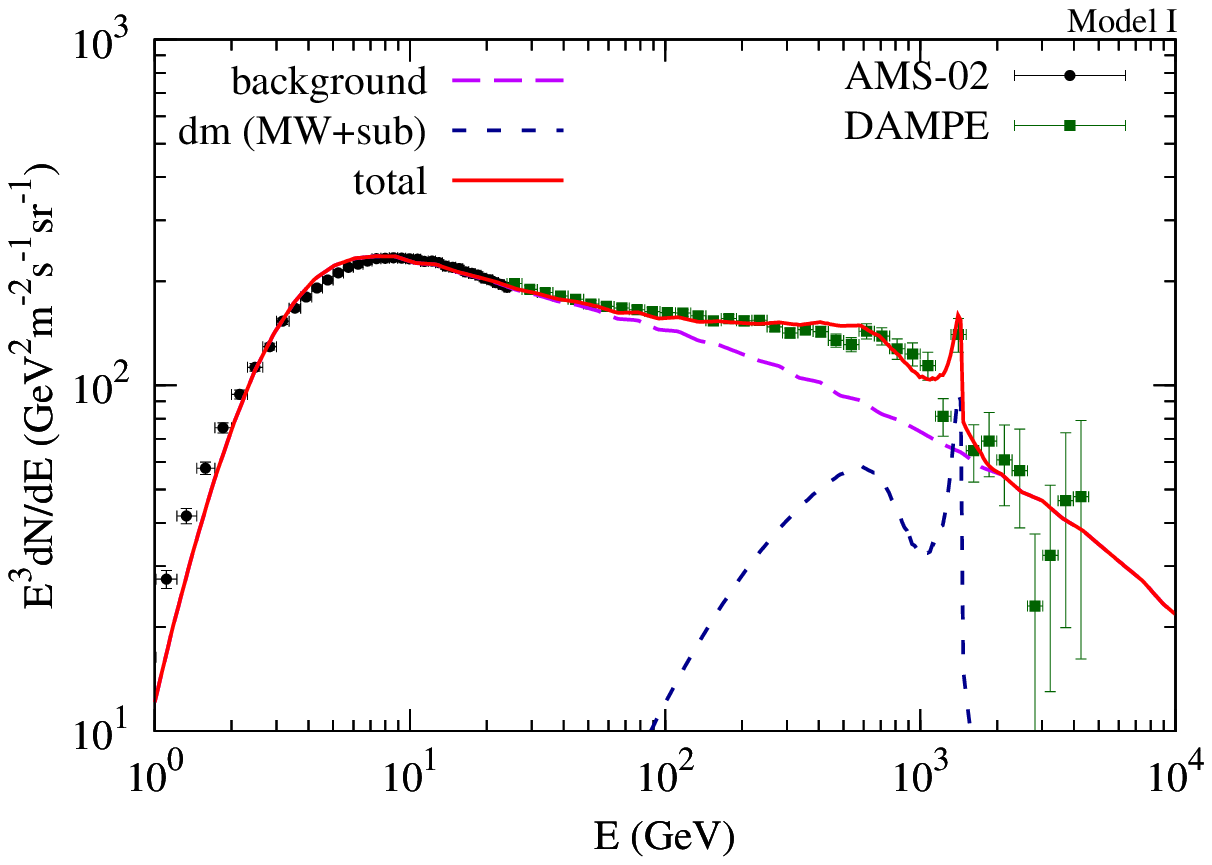}
\includegraphics[width=0.48\columnwidth]{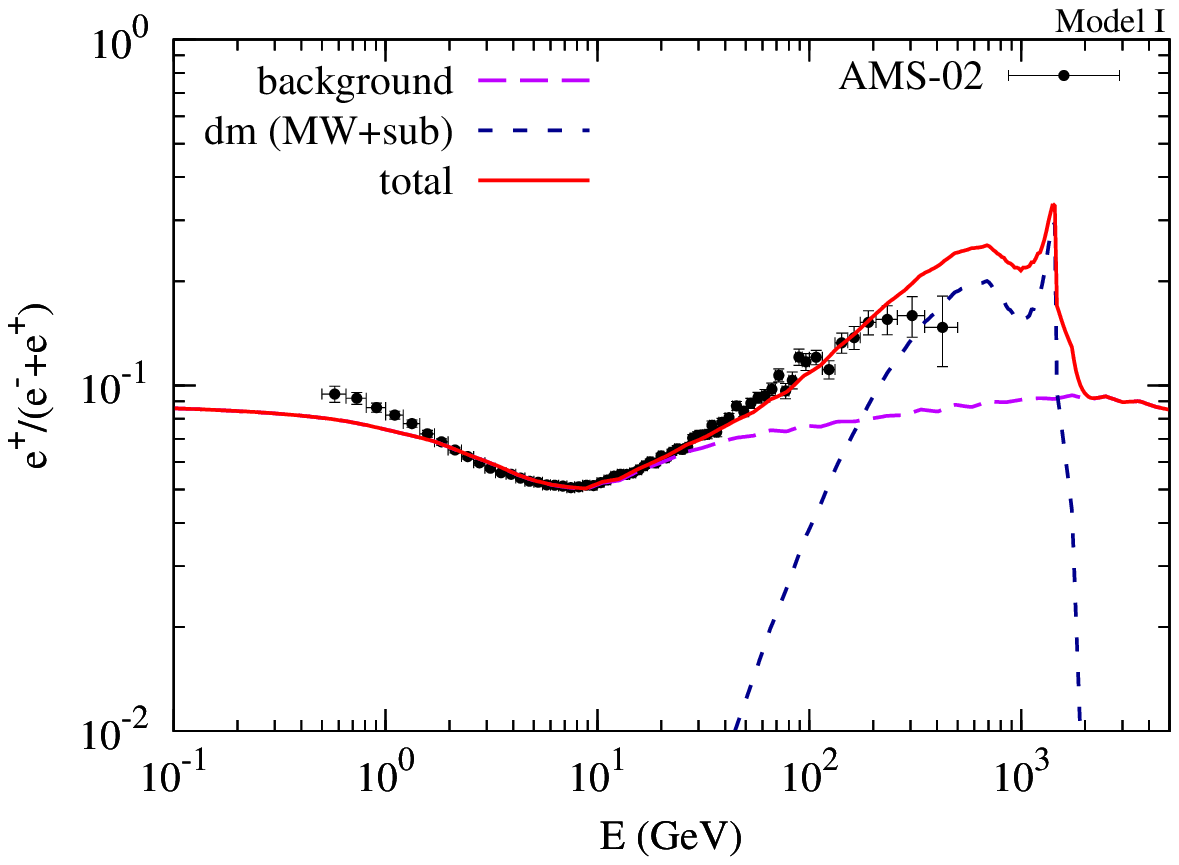}
\includegraphics[width=0.48\columnwidth]{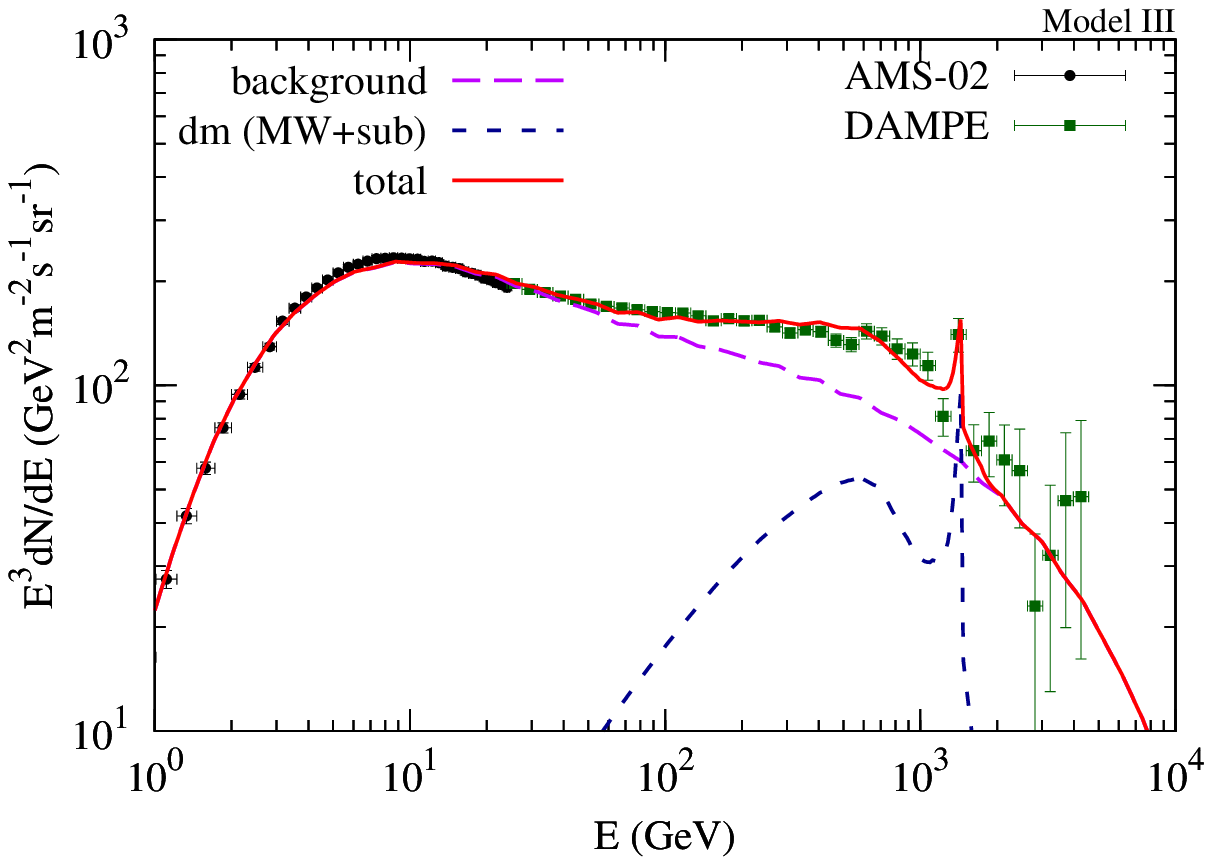}
\includegraphics[width=0.48\columnwidth]{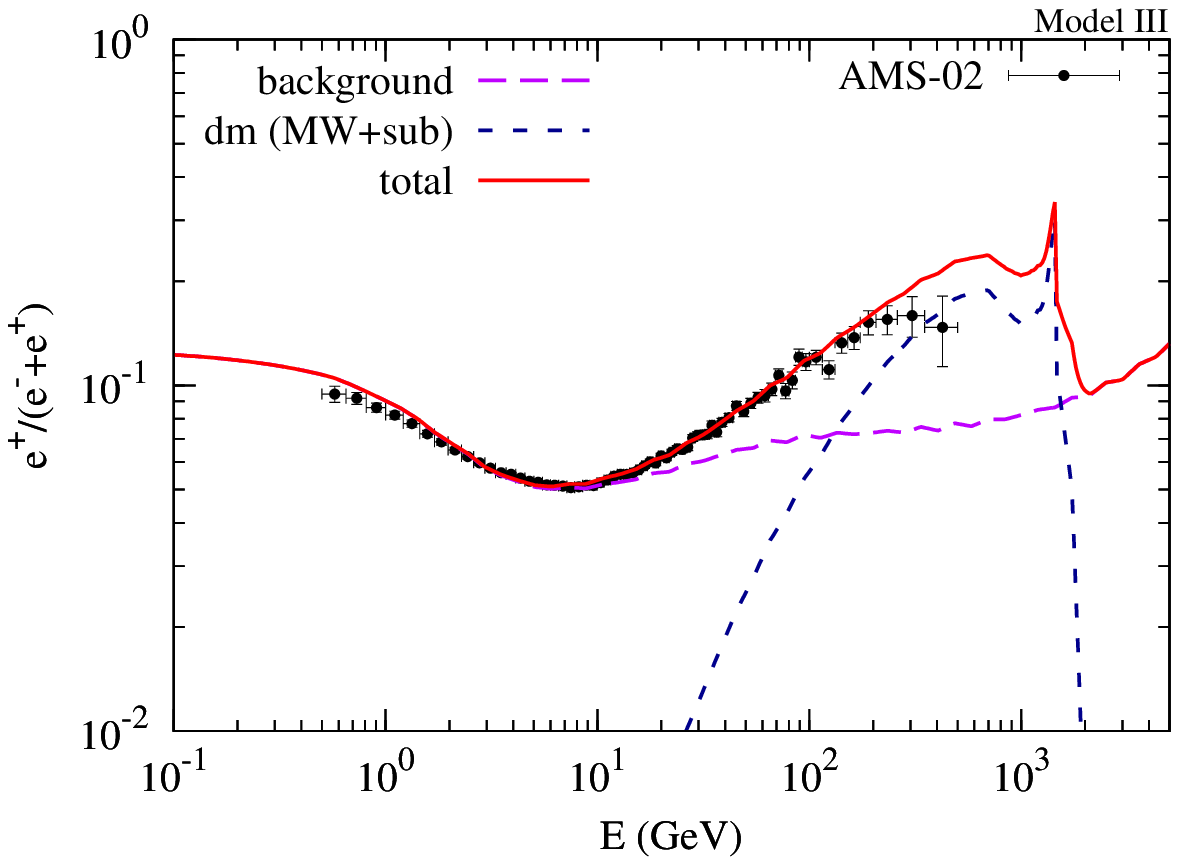}
\includegraphics[width=0.48\columnwidth]{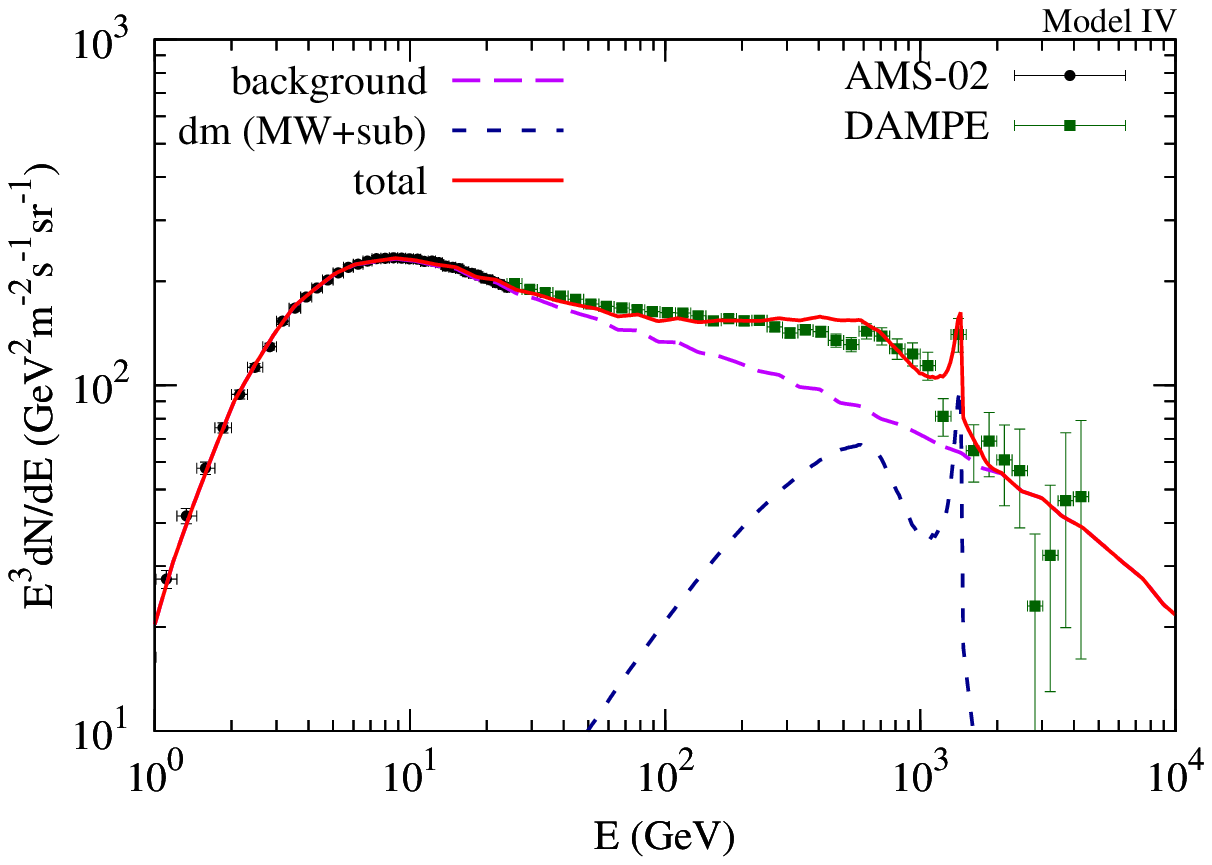}
\includegraphics[width=0.48\columnwidth]{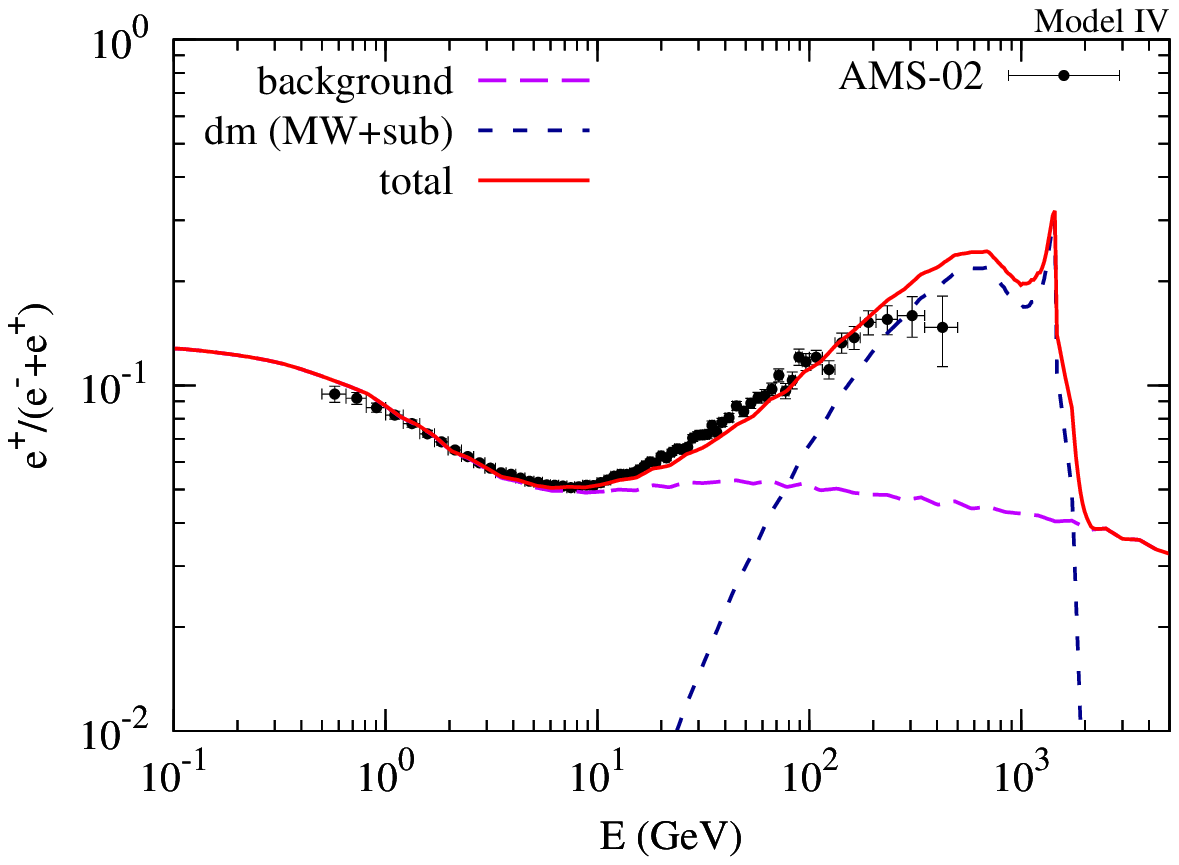}
\caption{Same as Figure~\ref{equalcontribution} but for cosmic ray
propagation parameter settings I (top panels), III (middle panels),
and IV (bottom panels).}
\label{other}
\end{figure}
The comparisons between the model and data
for the other three propagations models are given in Figure \ref{other} for DM annihilation/decay branching ratios $e:\mu=1:1$. From the figure, we can see that these three propagation models match with the data very well.
Similar to CR propagation model II, the annihilation cross section and decaying lifetime as shown in in Table \ref{tab2} are all
consistent with the current $\gamma$-rays \cite{Chen:2009uq,
Hutsi:2010ai,Cirelli:2012ut,Cheng:2016slx} and CMB \cite{Ade:2015xua} observations.

\end{document}